# Resistance profile measurements on a symmetric electrical pulse induced resistance change device

X. Chen*, J. Strozier, N.J. Wu, A. Ignatiev

*Texas Center for Advanced materials,*

*University of Houston, Houston, TX 77204-5004*

We report the first direct measurements of the micro scale resistance profile between the terminals of a two terminal symmetric thin film $Pr_{0.7}Ca_{0.3}MnO_3$ electrical pulse induced resistance change device composed of a $Pr_{0.7}Ca_{0.3}MnO_3$ active layer. The symmetric device is one in which the electrode shape, size, composition, and deposition processing are identical. We show that under certain limitations of pulse switching voltage, such a symmetric electrical pulse induced resistance change device can exhibit either no net device resistance switching at room temperature, or bipolar switching with the resistance hysteresis curve exhibiting a "table leg" structure. The resistance measurements are made using surface scanning Kelvin probe microscopy, which allows for the measurement of the profile of resistance from one electrode, across the $Pr_{0.7}Ca_{0.3}MnO_3$ material and into the second electrode, both before resistance switching and after switching. The results show that resistance switching in the symmetric device occurs primarily in the interface region within about 1 to 3 micron of the electrical contact surface. Resistance switching is also observed in the bulk $Pr_{0.7}Ca_{0.3}MnO_3$ material although at a lower level. Symmetry considerations for a two terminal symmetric device that can switch resistance are discussed, and the data reported here is consistent with the symmetric model previously developed.

[*] Email: xinchen@svec.uh.edu; Fax: 713-747-7724

Interactions among charges, electronic spins and orbital orderings account for a rich variety of behavior in transition-metal oxides.[1] Among them, the electrical pulse induced resistance change (EPIR) effect is one of the more interesting phenomena discovered recently in perovskite oxides with special attention to colossal magnetoresistance (CMR) oxides at room temperatures.[1-6] An EPIR device is a room temperature, non-volatile, reversible resistance switching device which is switched by the application of short electrical pulses across two electrodes[2]. It is the basis for a resistive random access memory (RRAM)[3], which has the projected advantage of non-volatility, fast programming, small bit cell size and low power consumption. Studying the EPIR effect is not only important for understanding the physics of the EPIR effect, and the importance of the CMR effect in its operation, but is also critical for advancing high density RRAM that can overcome shortcomings in current semiconductor memories.[5] Currently, however, there are questions as to whether or not the EPIR mechanism is a bulk[2,7] or interface effect.[8] Underlying such concerns is a more profound question of symmetry of the device. A symmetric EPIR device is one in which both electrodes are of the same material and have the same contact area with the bulk material (PCMO). It has been reported[8] that the "symmetric" device appears to violate conservation of parity. That is, it seems that the bulk PCMO material switches its resistance, a scalar, either up or down with respect to the polarity of the switching pulse, a vector. To explain this apparent parity violation, arguments are given in the literature that include surface/interface effects in which the switching material next to each electrode/material interface reacts to the polarity of the switching pulse in a different way, therefore removing the symmetry. Possibilities are: poor sample quality, different material used for the two electrodes, making devices with only one functioning interface between metal electrode and the active perovskite material, which might result in asymmetric oxidation, asymmetric concentration of cations, diffusion of oxygen vacancies, asymmetric trapped charge carriers, phase separation, etc.[8-10] Without invoking those arguments all based on some intrinsic asymmetry, we will show in this communication, that a symmetric device as previously defined, generally shows no net resistance switching behavior, but can indeed switch resistance states provided that the switching pulse voltage for at least one polarity is kept within certain limits. Our results also show that resistance switching in our symmetric device occurs primarily in the interface region within ~1-3μm of the electrical contact surface. A resistance switching is also observed in the remainder the active PCMO material although at a lower level.

An EPIR device was fabricated using a $Pa_{0.7}Ca_{0.3}MnO_3$ (PCMO) thin film as the active layer with silver electrodes of ~ 25 μm separation deposited on top of the PCMO film to form the device. The surface microstructure of the device was measured by AFM, and scanning Kelvin probe microscopy (SKPM), based on the capacitance change between a vibrating tip and the sample, was used to probe the sample surface potential.[11, 12] Such potential measurement can be used to get sample resistance information.[13] Fig. 1A shows the sample schematic and a conducting SPM tip scanning over the Ag-PCMO-Ag device surface to detect the topography and potential distribution across the device, which lies between the two silver electrodes B and C. A current source and voltage meter are connected to the sample for the electrical measurements, and a pulse generator is used to apply pulses to the device across electrodes B and C for resistance switching.

The device topography as measured by AFM is shown in Fig. 1B, where the higher bright regions at the two ends are from the two silver electrodes. The dark brown lower area is from the PCMO epitaxial thin film of the device, which is generally uniform with some nano-sized particles scattered at random. The RMS roughness of the PCMO film is ~5nm.

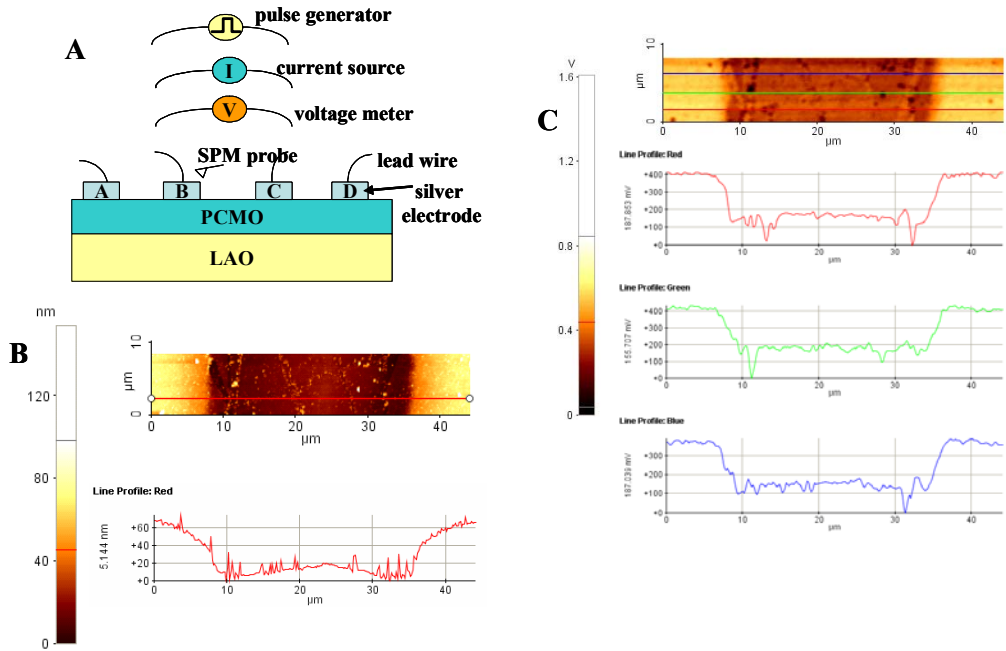

**Fig. 1** The scanning Kelvin probe microscopy (SKPM) experiment set up and measurements on the Ag-PCMO-Ag switching device. (A) a schematic diagram for SKPM experiment; (B) an AFM surface scan and line scan of the device structure; (C) the SKPM 2D potential plot and three separate line scans acquired at the same time as the AFM scan under zero bias current.

The surface potential map shown in Fig. 1C was obtained via SKPM taken at the same time as the AFM scan. It is a sharp image showing a surface potential distribution correlated with the device structure: bright yellow high potential areas consistent with the silver electrodes, brown low potential area of the PCMO thin film between the electrodes, and potential blips from randomly scattered nanoparticles on the PCMO surface, and /or the effects of the photo-lithographic patterning of the Ag electrodes on the PCMO. Three different potential line scans shown in Fig 1C are corresponding to three different positions on the sample which are color coded with the SKPM surface potential map in Fig. 1C.

An approximately 200mV potential step across the Ag/PCMO interfaces is also seen in Fig 1C. Since PCMO is a p-type semiconductor in contact with a silver metal electrode, the resulting potential jump is consistent with a metal/p-type semiconductor Schottky-like barrier in which the PCMO is negatively charged with respect to the silver electrode.

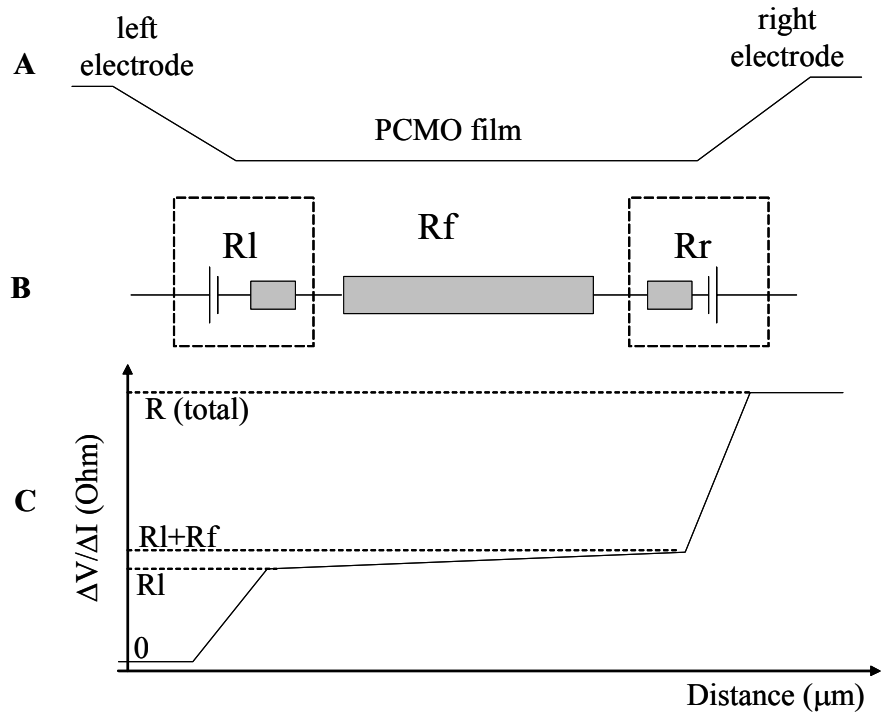

**Fig. 2** A schematic for the SKPM resistance measurement on the Ag-PCMO-Ag sample. (A) the schematic sample cross section and the equivalent circuit model for the sample; (B) the expected ΔV/ΔI vs distance curve from SKPM from which resistance distribution in the device can be read out directly

Fig. 2A shows the equivalent circuit for our symmetrical device as related to the sample cross section. Rl is the resistance of the left interface region, Rr the resistance of the right interface region and Rf the resistance of the PCMO film. The voltage sources at Rl and Rr are the open circuit contact potentials.

The presence of a potential barrier must be accounted for in any resistance measurements as the resistance of the sample is not simply the applied voltage plus the barrier height divided by the current. Instead we measure the resistance at any point relative to ground as the value of the voltage change over the current change: i.e. $\Delta V/\Delta I$ for small values of V so that no switching occurs in the measurement process. In this way, the work function value at each point cancels out, yielding the resistance of the device internal to the surface.

A schematic resistance measurement from SKPM is shown in Fig. 2B. The left electrode is set to ground potential and we plot $\Delta V/\Delta I$ vs. distance. The value $\Delta V/\Delta I$ at any point on the curve represents the resistance from that point to the left electrode, and the $\Delta V/\Delta I$ difference between any two points measures the resistance between them. Therefore the value of the resistance in the PCMO film, the interfaces, and the total device can be directly read from these curves.

Fig. 3A, shows the potential map of the sample under small bias current (1μA) measuring conditions, and a voltage line scan across the sample. It should be noted that the two electrodes are not at the same potential, with the lighter color representing the higher positive potential (right side of sample). The resistance, $\Delta V/\Delta I$, as a function of distance is plotted in Fig. 3B. The total resistance across the sample is about 200kohms. Note that about 98% of the total resistance occurs over regions of about 1-3 μm near the interfaces at each of the two electrodes. Close examination of Fig. 3B indicates that ~2% of the resistance is from the PCMO thin film material away from the interface regions.

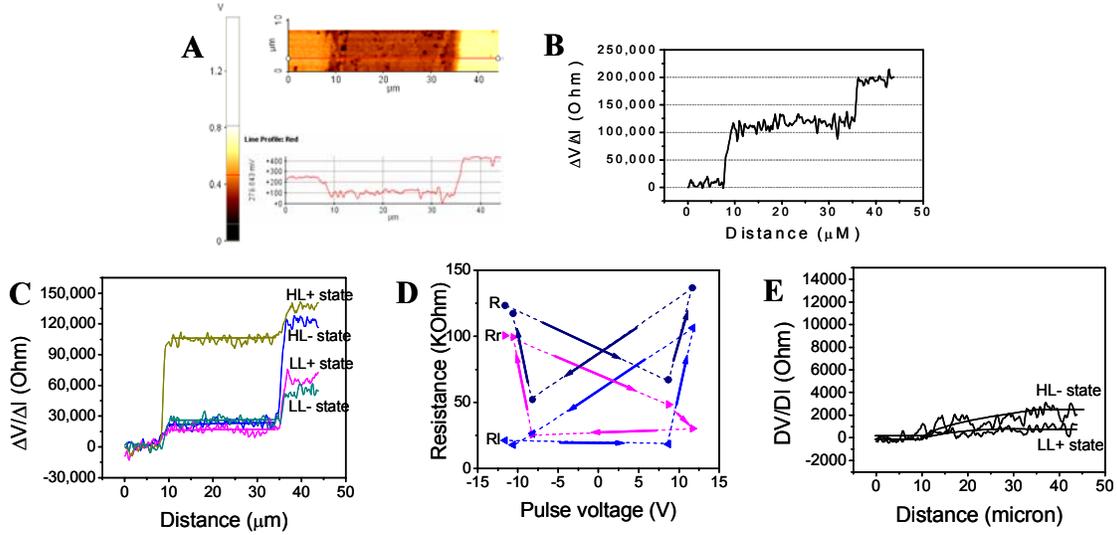

**Fig. 3** Direct SKPM observation of resistance switching in the Ag-PCMO-Ag device. (A) the surface potential scan and a line scan under a current bias from right to left; (B) the resultant ΔV/ΔI curve from the line scan; (C) sample ΔV/ΔI curves after pulses of varying amplitude and polarity have been applied to the sample yielding different resistance states for the sample; (D) device element resistances extracted from the resistance state curves of (C) showing a symmetric switching behavior (arrows and dash lines added to guide the eye); (E) SKPM measurement of the PCMO film region of the sampled under the HL- and LL+ switch conditions indicating PCMO resistance change.

The device was then subjected to 600ns pulses of various amplitudes of the order of 10 volts and of both polarities so as to resistively switch the device. In general, the measured resistance between the two electrodes B and C, can be characterized as four distinct states, and their integrated resistance as a function of distance is summarized in Fig. 3C. These four states can be characterized by the resistance values of the 3 resistances in the model given in Fig. 2A. They are: HL- for low $R_l$, high $R_f$ and high $R_r$; HL+ for high $R_l$, low $R_f$, and low $R_r$; and LL+ and LL- when $R_l$, $R_f$, and $R_r$ are low. The meaning of the "+" and "-" denotes that the state resulted from a positive or a negative pulse. Differences in $R_l$ high and $R_r$ high, and $R_l$ low and $R_r$ low give rise to the 4 distinct states as noted above. When the device is first subjected to a saturation voltage pulse of -12V or greater and 600 ns (HL- state), we find that the value of $R_r$ remains close to its original value of about 100kohms, while Rl switches to a much lower (28kohms) value, and Rf increases to about 2kohms. If the same pulse (duration and absolute magnitude) of now opposite polarity (+12V) is applied (HL+ state), the value of Rl is restored to its previous value of about 100kohms, while Rr is switched down to about 28kohms, and Rf decreases by about 2kohms.

It is well to note that the total resistance of the device in the HL- state and the HL+ state is about the same, so there is essentially no net device switching although individual regions of the device switched. In fact, we can define a device to be exactly symmetric if it exhibits no overall change in its total resistance due to switching pulses of magnitude greater than a saturation voltage (in this case about ±12V). On the other hand, overall switching can occur in a symmetric device if the voltage of one of the switching pulses is in a range less than the saturation value.

The resistance of these switching states as a function of distance is also plotted in Fig. 3C. Starting with the device in the HL- state, an application of a +10V pulse (at 600 ns) causes the resistance of Rr to drop but the pulse is insufficient to cause the Rl resistance to switch up. This results in the LL+ state where both Rl and Rr are at their low resistances. The "+" denotes that the LL+ state results from a positive pulse applied to the HL- state. If the pulse amplitude is increased to +12V (the saturation voltage) or larger, then Rl switches up and the device is in the saturated HL+ state.

Starting with the device in the HL+ state, application of a -10V pulse causes the resistance of the Rl to drop, but the pulse is insufficient to cause the Rr resistance to switch up. This results in the LL- where both Rl and Rr are at their low resistances. The "-" denotes that the LL- state results from a negative pulse applied to the HL+ state. If the pulse amplitude is increased (in absolute magnitude) to -12V (the saturation voltage) or larger, then Rr switches up and the device is in the saturated HL- state.

In both these cases (LL+ and LL-), it is observed that the high resistance interface (Rl or Rr) decreases at the lower pulse voltage. The low resistance interface switches up when the pulse voltage increases to the saturation value. Fig. 3D shows the resistance switching behavior of R, Rl and Rr with pulsing starting from –12V to +12V, and back to –12V (the arrows are added to guide the eye).

The PCMO film resistance switching is further checked with SKPM, with the results shown in Fig. 3E. Current is applied to the two outer electrodes (A, D in Fig. 1A) with SKPM performed between the two inner electrodes (B, C in Fig. 1A). A new tip was used and the current was increased to 5μA so that the measurement noise was reduced from ~10kohm in Fig. 3C to <~1kohm. Fig. 3E shows that there is a PCMO film resistance, Rf, switch of about 2kohms between the HL- state and the LL+ state.

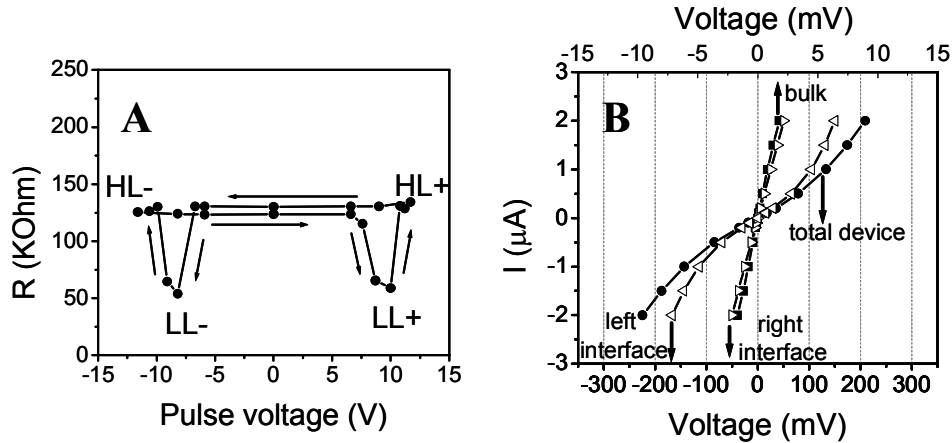

**Fig. 4** Four point measurement of resistance supporting the SKPM switching data. (A) the symmetric device R vs pulse voltage hysteresis loop showing the "table leg" structure; (B) the I-V curves of the device resistance elements with the device in the HL+ state.

The partial switching results for the LL+ and LL- states are supported by four point resistance measurements shown in Fig. 4A, where the total sample resistance is plotted versus the pulse voltage. Without saturation in the low resistance states, the symmetric device shows a large R switch ratio between HL- and LL+ and HL+ and LL- of about 100% accompanied by the "table leg" structure of the hysteresis curve the ends of which show no device switching under saturation pulse application.

I-V curves were also taken for each section: the left interface, right interface, and the PCMO film for each of the 4 resistance states (HL-, LL+, HL+, and LL-). A typical set of I-V curves (HL+ state) are presented in Fig. 4B. From this data, several observations can be made: i.) the low resistance interfaces and the very low resistance PCMO film yield I-V curves that are linear (or ohmic) over the current range -3µA to +3µA. ii.) the high resistance interfaces yield I-V curves that are linear for small values of the current (less than 1µA), but become both not linear and antisymmetric at higher values of the current. These interface switching results agree with that observed by others[6, 14] – in general the low resistance state tends to be close to ohmic, whereas the high resistance state exhibits non-linearity and asymmetry at the higher values of current.

The observed switching of a symmetric device is important because it has been previously suggested[8] that a symmetric device can not switch, as switching of such a electronic device would constitute a violation of parity conservation. However, in the actual symmetric device developed in this work, switching does occur. We have recently developed a phenomenalogical model,[15] which well describes this observed switching behavior. Furthermore, our modeling and experiments show such a symmetric device exhibits a pulse polarity dependent switching if the pulsing voltage amplitude is asymmetric, i.e., full saturation at one polarity and within certain limits less than saturation for the opposite polarity.

We also report that while the majority of the switching occurs in the interface regions, there is some switching in the PCMO film. If we consider the possibility that the PCMO film is made up of a large number of small regions of PCMO due to possible phase separation as reported by others,[16-18] each of the regions could switch similarly to the large symmetric device reported here. PCMO film switching could thus be explained as resistive switching of a large number of smaller units connected in a network. This explanation might point to bulk switching being a current rather than voltage-driven mechanism since the voltage drop across each individual unit would be very small, but current density would be about the same. Considering the switching conditions in our Ag-PCMO-Ag device, the pulse voltage drop across the PCMO film is estimated to be less than ~0.2V. However, if we calculate the current, the pulse intensity is estimated to be $>10^3 A/cm^2$ through the PCMO film area, and such high current density may generate a current migration effect in the film.[19] Further study is underway.

We have fabricated a symmetric EPIR device which when exposed to short saturation voltage pulses, shows no net device resistance switching. However, when exposed to less than saturation voltage pulses, shows significant EPIR net device resistance switching. Using SKPM, we have developed a method to directly observe the resistance and resistance change at a microscopic level across a symmetric EPIR device, and have measured the resistance switching at each interface and in the PCMO film of the device. We find that about 98% of the resistance switching occurs in the interface regions within a width of about 1-3 μm of the electrical contact surface, and ~2% resistance change occurs in the PCMO film.

The experimental data reported here also agree with the analysis of our symmetric model proposed earlier.[15] In particular, switching is possible in a symmetric device if the voltage amplitude of one of the two switching pulses of opposite polarities is in some range (depending on the material) that is less than the saturation switching voltage.

**Acknowledgements**

The assistance from A. Zomorrodian and L. Smith is very much appreciated. This research was partially supported by NASA, the State of Texas through TcSAM, Sharp Laboratories of America, and the R. A. Welch Foundation.